\documentclass{svmult}

  \def\pp{{\mathchoice
          {
              \kern 1pt%
              \raise 1pt
              \vbox{\hrule width5pt height0.4pt depth0pt
                    \kern -2pt
                    \hbox{\kern 2.3pt
                          \vrule width0.4pt height6pt depth0pt
                          }
                    \kern -2pt
                    \hrule width5pt height0.4pt depth0pt}%
                    \kern 1pt
           }
            {
              \kern 1pt%
              \raise 1pt
              \vbox{\hrule width4.3pt height0.4pt depth0pt
                    \kern -1.8pt
                    \hbox{\kern 1.95pt
                          \vrule width0.4pt height5.4pt depth0pt
                          }
                    \kern -1.8pt
                    \hrule width4.3pt height0.4pt depth0pt}%
                    \kern 1pt
            }
            {
              \kern 0.5pt%
              \raise 1pt
              \vbox{\hrule width4.0pt height0.3pt depth0pt
                    \kern -1.9pt  
                    \hbox{\kern 1.85pt
                          \vrule width0.3pt height5.7pt depth0pt
                          }
                    \kern -1.9pt
                    \hrule width4.0pt height0.3pt depth0pt}%
                    \kern 0.5pt
            }
            {
              \kern 0.5pt%
              \raise 1pt
              \vbox{\hrule width3.6pt height0.3pt depth0pt
                    \kern -1.5pt
                    \hbox{\kern 1.65pt
                          \vrule width0.3pt height4.5pt depth0pt
                          }
                    \kern -1.5pt
                    \hrule width3.6pt height0.3pt depth0pt}%
                    \kern 0.5pt
            }
        }}

  \def\mm{{\mathchoice
                       {
                             \kern 1pt
               \raise 1pt    \vbox{\hrule width5pt height0.4pt depth0pt
                                  \kern 2pt
                                  \hrule width5pt height0.4pt depth0pt}
                             \kern 1pt}
                       {
                            \kern 1pt
               \raise 1pt \vbox{\hrule width4.3pt height0.4pt depth0pt
                                  \kern 1.8pt
                                  \hrule width4.3pt height0.4pt depth0pt}
                             \kern 1pt}
                       {
                            \kern 0.5pt
               \raise 1pt
                            \vbox{\hrule width4.0pt height0.3pt depth0pt
                                  \kern 1.9pt
                                  \hrule width4.0pt height0.3pt depth0pt}
                            \kern 1pt}
                       {
                           \kern 0.5pt
             \raise 1pt  \vbox{\hrule width3.6pt height0.3pt depth0pt
                                  \kern 1.5pt
                                  \hrule width3.6pt height0.3pt depth0pt}
                           \kern 0.5pt}
                       }}

\catcode`@=11
\def\un#1{\relax\ifmmode\@@underline#1\else
        $\@@underline{\hbox{#1}}$\relax\fi}
\catcode`@=12

\let\du=\du                     

\def\a{\alpha}
\def\b{\beta}
\def\c{\chi}
\def\d{\delta}

\def\f{\phi}
\def\g{\gamma}
\def\h{\eta}

\def\j{\psi}
\def\k{\kappa}

\def\m{\mu}
\def\n{\nu}

\def\q{\theta}

\def\s{\sigma}

\def\x{\xi}

\def\F{\Phi}
\def\G{\Gamma}

\def\ve{\varepsilon}

\def\bo{{\raise-.5ex\hbox{\large$\triangle$}}}          

\def\pa{\partial}                                       
\def\de{\nabla}                                         
\def\TH{{\raise.2ex\hbox{$\displaystyle \bigodot$}\mskip-4.7mu \llap H \;}}
\def\face{{\raise.2ex\hbox{$\displaystyle \bigodot$}\mskip-2.2mu \llap {$\ddot
        \smile$}}}                                      

\def\Tilde#1{\widetilde{#1}}                    
\def\Bar#1{\overline{#1}}                       
\def\leftrightarrowfill{$\mathsurround=0pt \mathord\leftarrow \mkern-6mu
        \cleaders\hbox{$\mkern-2mu \mathord- \mkern-2mu$}\hfill
        \mkern-6mu \mathord\rightarrow$}
\def\dvec#1{\vbox{\ialign{##\crcr
        \leftrightarrowfill\crcr\noalign{\kern-1pt\nointerlineskip}
        $\hfil\displaystyle{#1}\hfil$\crcr}}}           
\def\dt#1{{\buildrel {\hbox{\LARGE .}} \over {#1}}}     

\def\frac#1#2{{\textstyle{#1\over\vphantom2\smash{\raise.20ex
        \hbox{$\scriptstyle{#2}$}}}}}                   
\def\sfrac#1#2{{\vphantom1\smash{\lower.5ex\hbox{\small$#1$}}\over
        \vphantom1\smash{\raise.4ex\hbox{\small$#2$}}}} 
\def\bfrac#1#2{{\vphantom1\smash{\lower.5ex\hbox{$#1$}}\over
        \vphantom1\smash{\raise.3ex\hbox{$#2$}}}}       
\def\afrac#1#2{{\vphantom1\smash{\lower.5ex\hbox{$#1$}}\over#2}}    
\def\on#1#2{\mathop{\null#2}\limits^{#1}}               
\def\bvec#1{\on\leftarrow{#1}}                  

\def\[{\lfloor{\hskip 0.35pt}\!\!\!\lceil}
\def\]{\rfloor{\hskip 0.35pt}\!\!\!\rceil}

\def\du#1#2{_{#1}{}^{#2}}
\def\ud#1#2{^{#1}{}_{#2}}

\def\fracm#1#2{\hbox{\large{${\frac{{#1}}{{#2}}}$}}}
\def\ha{{\fracmm12}}

\def\un{\underline}
\def\fracmm#1#2{{{#1}\over{#2}}}

\def\low#1{{\raise -3pt\hbox{${\hskip 0.75pt}\!_{#1}$}}}

\def\Dot#1{\buildrel{_{_{\hskip 0.01in}\bullet}}\over{#1}}
\def\dt#1{\Dot{#1}}

\def\Tilde#1{{\widetilde{#1}}\hskip 0.015in}

\newskip\humongous \humongous=0pt plus 1000pt minus 1000pt
\def\caja{\mathsurround=0pt}
\def\eqalign#1{\,\vcenter{\openup2\jot \caja
        \ialign{\strut \hfil$\displaystyle{##}$&$
        \displaystyle{{}##}$\hfil\crcr#1\crcr}}\,}
\newif\ifdtup

\newcommand{\be}{\begin{equation}}
\newcommand{\ee}{\end{equation}}
\newcommand{\nbe}{\begin{equation*}}
\newcommand{\nee}{\end{equation*}}

\begin{document}
\title{Non-anti-commutative Deformation of Complex Geometry}
\author{ Sergei V. Ketov}
\institute{Department of Physics, Tokyo Metropolitan University, Hachioji-shi,
Tokyo 192--0397, Japan; {\sl ketov@phys.metro-u.ac.jp}}
\maketitle

\begin{abstract}
In this talk I review the well known relation existing between extended 
supersymmetry and complex geometry in the non-linear sigma-models, and then 
briefly discuss some recent developments related to the introduction of the
non-anti-commutativity in the context of the supersymmetric non-linear 
sigma-models formulated in extended superspace. This contribution is suitable 
for both physicists and mathematicians interesting in the interplay between 
geometry, supersymmetry and non(anti)commutativity. 
\end{abstract}

\section{Introduction}

Being a theoretical physicist, one gets used to mathematical tools, whose role
in modern theoretical high-energy physics is indispensable and indisputable. 
Especially when the experimental base is limited or does not exist, the use of 
advanced mathematics to get new insights in physics is particularly popular.
So it is no surprise that mathematical knowledge of theoretical physicists
is quite high. However, the way of dealing with mathematics in theoretical 
physics is different from that commonly used by mathematicians, and even the 
motivation and goals are different, as is also well known. I would like to
draw attention to another, less known fact that physical considerations may 
sometimes lead to new mathematics, or rediscovery of some famous mathematical
facts. In my contribution to this workshop, aimed towards more cooperation and
understanding between physicists and mathematicians, I would like to explain 
how investigation of supersymmetry in field theory of the non-linear 
sigma-models might have led to rediscovery of complex geometry and related 
mathematical tools. In addition, I would like to explain how introducing more 
fundamental structure (namely, non-anti-commutativity in superspace) leads to 
a new deformation of complex geometry, whose geometrical significance is yet
to be understood.

The paper is organized as follows. In sect.2 the basic notions of the
non-linear sigma-models are introduced. My presentation is `minimal' on
purpose, without going into details and/or many generalizations that might be 
easily added. I just summarize the basic ideas. In sect.~3, I introduce a
simple superspace, and review the known relation between extended 
supersymmetry and complex geometry in the non-linear sigma-models, by getting
all basic notions of complex geometry from a single and straightforward 
field-theoretical calculation. In sect.~4 some extended superspace techniques, 
making extended supersymmetry to be manifest, are briefly discussed.  In 
sect.~5 some more superspace structure is added by introducing 
the notion of {\it Non-Anti-Commutativity} (NAC) or `quantum superspace', and
its impact on the non-linear sigma-model target space is calculated. The 
simplest non-trivial explicit example of the NAC-deformed $CP(1)$ metric is 
given in  sect.~6. Our conclusion is sect.~7.
   
\section{Non-linear sigma-models}

The {\it Non-Linear Sigma-Model} (NLSM) is a scalar field theory whose 
(multi-component) scalar field $\f^a(x^{\m})$ is defined in a $d$-dimensional 
`spacetime' or a 'worldvolume' parametrized by local coordinates $\{x^{\m}\}$,
$\m=1,2,\ldots,d$. The fields $\f^a$ take their values in a $D$-dimensional 
Riemannian manifold $M$, called the NLSM target space, $a=1,2,\ldots,D$. The 
NLSM field values $\f^a$ can thus be considered as a set of (local) 
coordinates in $M$, whose metric is field-dependent. The NLSM format is the 
very general
field-theoretical concept whose geometrical nature is the main reason for many
useful applications of NLSM in field theory, string theory,  condensed matter 
physics and mathematics (see e.g., the book \cite{nlsm} for much more).

We assume the NLSM spacetime or worldvolume to be flat Euclidean space $R^d$, 
for simplicity, so that the NLSM action is supposed to be invariant under
translations (with generators $P_{\m}$) and rotations (with generators 
$M_{\m\n}$) in $R^d$. Let $ds^2=g_{ab}(\f)d\f^ad\f^b$ be a metric in $M$. Then
a generic NLSM action is given by 
\be\label{action}
S_{\rm bos.}\[\f]= \int d^dx\,L(\pa_{\m}\f,\f)~,\qquad
L=\ha g_{ab}(\f)\d^{\m\n}\pa_{\m}\f^a\pa_{\n}\f^b + m^2 V(\f)~,\ee
where summation over repeated indices is always implied. The function $V(\f)$
is called a scalar potential in field theory with a mass parameter $m$. 
The higher derivatives
of the field $\f$ are not allowed in the Lagrangian $L$, with the notable 
exception of $d=2$ where an extra (Wess-Zumino) term may be added to 
eq.~(\ref{action}):
\be \label{WZ}
L_{\rm WZ} = \ha b_{ab}(\f)\ve^{\m\n}\pa_{\m}\f^a\pa_{\n}\f^b~.\ee
The 2-form $B=b_{ab}(\f)d\f^a\wedge d\f^b$ in eq.~(\ref{WZ}), is called a 
torsion potential in $M$, by the reason to be explained in the next sect.~1.3.
 In string theory, it is called a $B$-field (or a Kalb-Ramond field).

\section{Supersymmetric NLSM}

There are two different ways to supersymmetrize the NLSM: either in the
worldvolume, or in the target space. Here we only discuss the 
worldvolume supersymmetrization of NLSM, in the case of even $d$.~\footnote{In
 string theory, the world-sheet supersymmetrization is known as the
 {\it Neveu-Schwarz-Ramond} (NSR) approach, whereas the target space 
supersymmetrization is called the {\it Green-Schwarz} (GS) approach.} 
Then adding
supersymmetry amounts to the extension of the Euclidean space motion group
$SO(d)\times T^d$ to a supergroup, with the  key superalgebra relation
\be \label{susyalg}
 \{ Q^i\low{\a},\bar{Q}_{\dt{\b}j} \}_+ = 
2\s^{\m}_{\a\dt{\b}}P_{\m}\d\ud{i}{j}~, 
\ee
where the additional fermionic supercharges $Q$ and $\bar{Q}$ are chiral and 
anti-chiral spinors of $SO(d)$, repectively, in the fundamental
representation of the internal $U(N)$ symmetry group, denoted by latin indices 
$i,j=1,2,\ldots,N$. The $N$ here is a number of supersymmetries, so that the 
$N>1$ supersymmetry is called the extended one. The chiral $\s$-matrices in
eq.~(\ref{susyalg}) obey Clifford algebra in $d$ dimensions. 

As regards the NLSM, it is not difficult to demonstrate by using only 
group-theoretical arguments that $d\leq 6$ \cite{nlsm}, and when d=2 then 
$N\leq 4$ \cite{sev4}.

The model-independent technology for a construction of off-shell manifestly 
supersymmetric field theories is called superspace. To give an example, let
us consider the simplest case of the $N=1$ supersymmetric NLSM in $d=2$. The
two-dimensional complex coordinates $z$ and $\bar{z}$ can be extended by
the anti-commuting (Grassmann) fermionic (spinor) coordinates $\q$ and 
$\bar{\q}$ to form a superspace $(z,\bar{z},\q,\bar{\q})$. Tensor functions in 
superspace are called superfields. A superfield is always equivalent to a
supermultiplet of the usual fields, e.g.
\be\label{sfld}
 \F(z,\bar{z},\q,\bar{\q})=\f +\q\j + \bar{\q}\bar{\j} +\bar{\q}\q F~, \ee
in terms of the bosonic field  components $\f(z,\bar{z})$ and 
$F(z,\bar{z})$, and the fermionic field components 
$\j(z,\bar{z})$ and $\bar{\j}(z,\bar{z})$. 

The supercharges can be easily realized in superspace as the differential
operators
\be \label{scharges}
Q = \fracmm{\pa}{\pa\q} - \q\pa~,\qquad {\rm and}\qquad
\bar{Q} = \fracmm{\pa}{\pa\bar{\q}} - \bar{\q}\bar{\pa}~,\ee
where we have introduced the notation $\pa=\pa_z$ and 
$\bar{\pa}=\pa_{\bar{z}}$. It is not difficult to check that the covariant 
derivatives 
\be \label{covder}
D = \fracmm{\pa}{\pa\q} + \q\pa~,\quad
\bar{D} = \fracmm{\pa}{\pa\bar{\q}} +\bar{\q}\bar{\pa}~,\qquad \pa\quad
{\rm and}\quad  \bar{\pa}~,\ee 
all (anti)commute with the supercharges (\ref{scharges}) indeed, which allows 
us to use them freely in the covariant superspace action. Then the unique N=1
supersymmetric extension of the bosonic NLSM action (\ref{action}) is easily
constructed in N=1 superspace as follows (we ignore a scalar potential here):
\be \label{saction}
S_1 = \int d^2x d^2\q\, \left( g_{ab} + b_{ab}\right)D\F^a\bar{D}\F^b~,\ee
where both $g_{ab}(\F)$ and $b_{ab}(\F)$ are now functions of the superfields
$\F^a$ of eq.~(\ref{sfld}). The component fields $F$ appear to be 
non-propagating (they are called to be auxiliary), since they satisfy merely 
algebraic equations of motions. They are supposed to be substituted by 
solutions to their 'equations of motion'. Having evaluated  the Berezin 
integral in eq.~(\ref{saction}), one gets eq.~(\ref{action}) as the only purely
bosonic contribution that is modified by the fermionic terms, namely, by a
sum of the covariant Dirac term and the quartic fermionic interaction whose
field-dependent couplings are given by the curvature tensor with torsion.

The $B$-field, in fact, enters the field action $S_1$ only via its 
curl (= torsion in $M$)
\be \label{torsion}
T\ud{a}{bc}=-\fracm{3}{2}g^{ad}b_{[bc,d]}~,\ee
that, in its turn, enters the action only via the connections (in $M$)
\be\label{connections}
\G\ud{a}{\pm bc}=\left\{ \begin{array}{c} a \\ bc \end{array} \right\} 
\pm T\ud{a}{bc}~.
 \ee 

By construction the two-dimensional action $S_1$ is invariant under the N=1 
supersymmetry transformations
\be \label{susylaw}
\d_{\rm susy} \F^a=\ve Q \F^a +\bar{\ve}\bar{Q}\F^a~,\ee
with
\be \label{fermions}
\left. Q\F^a\right|=\j^a \quad {\rm and}\quad 
\left. \bar{Q}\F^a\right|=\bar{\j}^a~,\ee
where $|$ denotes the leading (i.e. $\q$- and $\bar{\q}$- independent) part of
a superfield, while $\ve$ and $\bar{\ve}$ are the infinitesimal 
fermionic (Grassmann) N=1 supersymmetry transformation parameters. 
Eq.~(\ref{fermions}) can serve as the definition of
the fermionic superpartners of the bosonic NLSM field $\f^a$.

It is not difficult to generalize the NLSM (\ref{saction}) by adding a 
scalar superpotential in superspace,
\be \label{1pot}
S_{\rm pot.} = m \int d^2x d^2\q\, W(\F)~,\ee
with a arbitrary real function $W(\F)$ and a mass parameter $m$. It gives rise
 to the scalar potential
\be \label{scapot} 
V(\f)=\ha m^2 g^{ab}(\f)\pa_a W(\f)\pa_b W(\f)~,\ee
while it does not modify the NLSM kinetic terms, as is already clear from 
dimensional reasons.

A generic two-dimensional NLSM with an arbitrary Riemannian target space $M$ 
(and no scalar potential) can always be N=1 supersymmetrized, as in 
eq.~(\ref{saction}). When $M$ is a Lie group manifold,
there is a preferred (group-invariant) choice for its metric and torsion, while
such NLSM is called a  {\it Wess-Zumino-Novikov-Witten} (WZNW) model 
\cite{nlsm}. 
One may also introduce the so-called {\it gauged} WZNW models with a 
homogeneous target space $G/H$, where $H$ is a subgroup of $G$. In
differential geometry, it corresponds to the quotient construction \cite{nlsm}.

The next relevant question is: which restrictions on the NLSM target space $M$,
in fact, imply more supersymmetry, i.e. $N>1$ ? To answer that question, all 
one needs is  to write down the most general Ansatz for the second
supersymmetry transformation law (it follows by dimensional reasons) in terms 
of the N=1 superfields as
\be\label{2susy}
\d_2 \F = \h J\ud{a}{b}(\F)D\F^b +\bar{\h}\bar{J}\ud{a}{b}(\F)\bar{D}\F^b~,
\ee
and then impose the invariance condition
\be \label{2inv}  \d_2S_1=0~.\ee
In equation (\ref{2susy}), the $\h$ and $\bar{\h}$ are the infinitesimal
parameters of the second supersymmetry, while $J\ud{a}{b}(\F)$ and  
$\bar{J}\ud{a}{b}(\F)$ are some tensor functions to be fixed by 
eq.~(\ref{2inv}). It is straightforward (though tedious) to check that the 
condition (\ref{2susy}) amounts to the following restrictions 
(see e.g., ref.~\cite{sevrin}):
\be \label{co-const}
 \de^+_cJ\ud{a}{b}=\de^-_c\bar{J}\ud{a}{b}=0~,\ee
and
\be \label{herm}
g_{bc}J\ud{c}{a}=-g_{ac}J\ud{c}{b}~,\qquad 
g_{bc}\bar{J}\ud{c}{a}=-g_{ac}\bar{J}\ud{c}{b}~.
\ee
In addition, one gets the standard (on-shell) N=2 supersymmetry algebra 
(\ref{susyalg}) provided that (see e.g., ref.~\cite{sevrin})
\be \label{comstr}
 J^2=\bar{J}^2=-{\bf 1} \quad {\rm and}\quad
N\ud{a}{bc}[J,J]=N\ud{a}{bc}[\bar{J},\bar{J}]=0~,\ee
where we have introduced the Nijenhuis tensor 
\be\label{nijen}
 N\ud{a}{bc}[A,B]=A\ud{d}{[b}B\ud{a}{c],d}+A\ud{a}{d}B\ud{d}{[b,c]}+
 B\ud{d}{[b}A\ud{a}{c],d}+ B\ud{a}{d}A\ud{d}{[b,c]}~.\ee

So we can already recognize (or re-discover) the basic notions of (almost) 
complex geometry, such as an (almost) complex structure, a hermitean 
metric, a covariantly constant (almost) complex structure, and an integrable 
 complex structure (see e.g., ref.~\cite{cdg}). To be precise, we get the
following theorem:

{\it a two-dimensional N=1 supersymmetric NLSM is actually (on-shell) N=2 
supersymmetric, if and only if (1) it allows two (almost) complex structures,
$J$ and $\bar{J}$, (2) the NLSM metric is hermitean with respect to each of 
them, and (3)  each (almost) complex structure is covarianlty constant with 
respect to the asociated $(\pm)$ connection, so that it is actually integrable.
}

The integrability here means the existence of holomoprhic and anti-holomorphic
coordinates (i.e. the holomorphic transitions functions) after rewriting a 
complex structure to the diagonal form (with the eigenvalues $i$ and $-i$).

It should be noticed that the complex structures $J$ and $\bar{J}$ may not
be commuting with each other, $\[J,\bar{J}\]\neq 0$, because they are 
covariantly constant with respect to the different connections in 
eq.~(\ref{connections}), respectively. For instance, the mixed N=2 
supersymmetry commutator  
\be \label{mixedcom}
\[ \d(\h),\d(\bar{\h})\]\F^a = 
 \h\bar{\h}\[J,\bar{J}\]\ud{a}{b}\left(D\bar{D}\F^b
+\G\ud{b}{-cd}D\F^d\bar{D}\F^c\right)~,\ee
is required to be vanishing by the N=2 supersymmetry algebra (\ref{susyalg}).
It is already true on-shell, i.e. when the NLSM equations of-motions,
$D\bar{D}\F^b+\G\ud{b}{-cd}D\F^d\bar{D}\F^c=0$, are satisfied, though it is 
also the case off-shell only if $\[J,\bar{J}\]=0$. The complex structures  $J$
 and $\bar{J}$ may not therefore be simultaneously integrable, in general. 
If, however they do commute, then the existence of an off-shell N=2 extended 
superspace formulation of such N=2 NLSM with manifest N=2 supersymmetry is 
guaranteed.

\section{NLSM in extended superspace}

To give the simplest example of the N=2 extended superspace in two dimensions
$(z,\bar{z})$, let's introduce two (Grassmann) fermionic coordinates for each
chirality, i.e. $(z,\q^+,\q^-)$ and $(\bar{z},\bar{\q}^+,\bar{\q}^-)$, and then
the N=2 supercharges $Q_{\pm}$ and $\bar{Q}_{\pm}$, the N=2 superspace 
covariant derivatives $D_{\pm}$ and $\bar{D}_{\pm}$, and N=2 scalar superfields
$\F^i(z,\bar{z},\q^+,\q^-,\bar{\q}^+,\bar{\q}^-)$, like in the N=1 case 
(see the previous section), where now $i=1,2,\ldots , m$.

However, there is the immediate problem: a general (unconstrained) N=2 scalar
superfield has a physical vector field component that is not suitable for the 
NLSM. The simplest way to remedy that problem is to use the (off-shell) N=2
chiral and anti-chiral superfields, subject to the constraints
\be \label{chiralsup}
\bar{D}_{\pm}\F=0 \qquad {\rm and}\qquad  D_{\pm}\bar{\F}=0~,\ee
respectively. Their most general NLSM action is then given by
\be \label{N=2action}
S_2 = \int d^2x d^2\q d^2\bar{\q} \, K(\F,\bar{\F}) +
 m\int d^2x d^2\q \, W(\F) + m\int d^2x d^2\bar{\q}\, \bar{W}(\bar{\F})~,\ee
in terms of a non-holomorphic kinetic potential $K(\F,\bar{\F})$ and a
holomorphic superpotential $ W(\F)$.  

A simple straightforward calculation of the NLSM metric out of 
eq.~(\ref{N=2action}) reveals a K\"ahler metric 
$g_{i\bar{j}}=\pa_i\bar{\pa}_jK$ with the K\"ahler potential $K$, and no 
torsion. Therefore, K\"ahler complex geometry could have been also 
re-discovered from the N=2 supersymmetric NLSM.

When one adds the so-called twisted chiral N=2 superfields, subject to the
following off-shell N=2 superspace constraints:
\be \label{twisted} 
\bar{D}_+\tilde{\F}=D_-\tilde{\F}=0 \quad {\rm and} \quad
\bar{D}_-\Bar{\tilde{\F}}=D_+\Bar{\tilde{\F}}=0~,\ee
their most general N=2 superspace action, 
\be \label{twistN=2action}
S_{2,T} = \int d^2x d^2\q d^2\bar{\q} \, K(\F,\bar{\F},\tilde{\F},
\Bar{\tilde{\F}}) +~{\rm obvious~superpotential~terms}~,\ee
would give rise to a non-trivial torsion too, though with the commuting
complex structures, $\[J,\bar{J}\]=0$ \cite{ghr}. Actually, the exchange
$\F\leftrightarrow \tilde{\F}$ corresponds to the T-duality in string theory. 

To get the most general N=2 supersymmetric NLSM with  $\[J,\bar{J}\]\neq 0$,
one has to add the so-called semi-chiral (reducible) N=2 superfields,
$\hat{\F}$ and $\Bar{\hat{\F}}$, subject to the off-shell N=2 superspace 
constraints \cite{bus,sevrin}
\be \label{semich}
\bar{D}_+\bar{D}_-\hat{\F}=0\qquad {\rm and}\qquad D_+D_-\Bar{\hat{\F}}=0~.\ee

When asking for even more supersymmetry in a two-dimensional supersymmetric 
NLSM, one gets three linearly independent (almost) complex structures of each
chirality, obeying a quaternionic algebra (see e.g., ref.~\cite{nlsm}),
\be \label{quater}
J_a^{\pm (A)b}J_b^{\pm (B)c}= -\d^{AB}\d^c_a +\ve^{ABC}J_a^{\pm (C)c}~,\quad 
{\rm where} \quad A,B,C=1,2,3~,\ee
and similarly for $\bar{J}$. They all must be covariantly constant,
\be \label{cc3}
\de^{\pm}J^{\pm}=0~,\ee
respectively. In particular, N=3 supersymmetry implies N=4 supersymmetry.

Unfortunately, a geometrical description of the two-dimensional N=4 
supersymmetric NLSM with torsion is still incomplete. In the case of the 
vanishing torsion, an N=1 supersymmetric NLSM is, in fact, N=4 supersymmetric
if and only if its target space is hyper-K\"ahler (see e.g., ref.~\cite{nlsm}
for more details). 

When a supersymmetric NLSM in question is, in fact, a 
(gauged) WZNW model, then its N=4 supersymmetry implies that its target space
must be a product of Wolf spaces \cite{gk,sevrin}. The Wolf space can be 
associated with any simple Lie group $G$. Let $\j$ be the highest weight root 
of $G$, and let $(E_{\j \pm},H)$ be the generators of the $su(2)_{\j}$ 
subalgebra of Lie algebra of $G$ (say, in Chevalley basis), associated with 
$\j$. Then the Wolf space is given by the coset
\be \label{wolf}
Wolf~space = \fracmm{G}{H_{\perp}\otimes SU(2)_{\j}}~,\ee
where we have introduced the $SU(2)_{\j}$ Lie group of the Lie algebra
 $su(2)_{\j}$ and the centralizer $H_{\perp}$ of the  $SU(2)_{\j}$ in $G$. 

An efficient off-shell N=4 superspace description of all two-dimensional
 N=4 supersymmetric NLSMs does not exist, though the use of harmonic superspace
\cite{hss} with the infinite number of auxiliary fields may be useful for
describing a large class of the manifstly N=4 supersymmetric NLSM.
 
It is also worth mentioning that the chiral generators of supersymmetry in two
 dimensions are independent, so that it is possible to have an unequal 
number of 'left' and `right' supersymmetries. It is called 'heterotic' or 
$(p,q)$ supersymmetry.~\footnote{It is conventional to set $p+q=2n$.} It is 
always possible to construct the minimal or $(1/2,0)$ supersymmetric
extention of any NLSM. A generic $(1/2,1)$ supersymmetric NLSM can be 
formulated in $(1/2,1)$ superspace. Less is known about other $(p,q)$ 
supersymmetric NLSM with $n=3,4$.

Finally, there is a simple relation between extended supersymmetry and higher
$(d>2)$ dimensions, which is just based on the representation theory of spinors
and Clifford algebras in various dimensions. A supersymmetric NLSM can
 first be formulated in six or four dimensions, and then it can be 
rewritten to lower dimensions, by simply restricting all its fields to be 
dependent upon lower number of their worldvolume coordinates (this procedure 
is called dimensional reduction). The manifestly supersymmetric formulation of
a higher-dimensional supersymmetric NLSM often requires the use of  
sophisticated (constrained) superfields \cite{nlsm}. In quantum theory, only 
two-dimensional NLSM are renormalizable, while their higher-dimensional 
counterparts are not. The same is true for the supersymmetric NLSM \cite{nlsm}.

\section{ Non-anticommutative deformation of four-dimensional 
supersymmetric NLSM} 

{\it Non-Anti-Commutativity} (NAC) or quantum superspace \cite{bs} is a
natural extension of the ordinary superspace, when the fermionic superspace
coordinates are assumed to obey a Clifford algebra instead of being Grassmann
(i.e. anti-commutative) variables \cite{kpt}. The non-anticommutativity 
naturally arises in the D3-brane superworldvolume, in the type-IIB constant 
Ramond-Ramond type background, in superstring theory \cite{ov}. In four 
dimensions, the NAC deformation is given by  
\be \label{nac} \{ \q^{\a},\q^{\b} \}_*=C^{\a\b}~~, \ee
where $C^{\a\b}$ can be identified with a constant self-dual gravi-photon
background \cite{ov}. The remaining N=1 superspace coordinates in 
the chiral basis ($y^{\m}=x^{\m}+i\q\s^{\m}\bar{\q}$, $\m,\n=1,2,3,4$ and 
$\a,\b,\ldots=1,2$) can still (anti)commute,
\be\label{nacss}
\[y^{\m},y^{\n}\]= \{ \bar{\q}^{\dt{\a}},\bar{\q}^{\dt{\b}} \}=
\{ \q^{\a},\bar{\q}^{\dt{\b}} \}=\[y^{\m},\q^{\a}\]=
\[y^{\m},\bar{\q}^{\dt{\a}}\]=0~.\ee
provided we begin with a four-dimensional {\it Euclidean}~\footnote{The 
Atiyah-Ward space-time of signature $(+,+,-,-)$ is also possible \cite{kgn}.} 
worldvolume having the coordinates $x^{\m}$. A supersymmetric field theory in 
the NAC superspace was extensively studied in the recent past, soon after the 
pioneering paper \cite{sei}. The choice (\ref{nacss}) preserves locality
in a NAC-deformed field theory.

The $C^{\a\b}\neq 0$ explicitly break the four-dimensional Euclidean 
invariance. The NAC nature of $\q$'s can be fully taken into account by using 
the (associative, but non-commutatvive) Moyal-Weyl-type (star) product of 
superfields,
\be \label{starp}
f(\q)* g(\q)=f(\q)\,
\exp\left(-\fracmm{C^{\a\b}}{2}\fracmm{\bvec{\pa}}{\pa
\q^{\a}}\fracmm{\pa}{\pa\q^{\b}}\right)g(\q)~, \ee
which respects the N=1 superspace chirality.~\footnote{We use the left 
derivatives as a default, the right ones are explicitly indicated.} The star 
product (\ref{starp}) is polynomial in the deformation parameter~,
\be \label{polstar}
 f(\q)*g(\q)=fg +(-1)^{{\rm deg}f}\fracmm{C^{\a\b}}{2}
\fracmm{\pa f}{\pa\q^{\a}}\fracmm{\pa g}{\pa\q^{\b}}-\det\,C
\fracmm{\pa^2 f}{\pa\q^2}\fracmm{\pa^2 g}{\pa\q^2}~~,\ee
where we have used the identity
\be \label{iden1}
\det C = \fracm{1}{2}\ve_{\a\g}\ve_{\b\d}C^{\a\b}C^{\g\d}~,\ee
and the notation
\be \label{not1}
\fracmm{\pa^2}{\pa\q^2}= \fracm{1}{4}\ve^{\a\b}\fracmm{\pa}{\pa\q^{\a}}
\fracmm{\pa}{\pa\q^{\b}}~~.\ee

We also use the following book-keeping notation for 2-component spinors:
\be \label{not2}
 \q\c=\q^{\a}\c_{\a}~,\quad  \bar{\q}\bar{\c}=\bar{\q}_{\dt{\a}}
\bar{\c}^{\dt{\a}}~,\quad \q^2= \q^{\a}\q_{\a}~,\quad 
\bar{\q}^2= \bar{\q}_{\dt{\a}}\bar{\q}^{\dt{\a}}.\ee
The spinorial indices are raised and lowered by the use of two-dimensional 
Levi-Civita symbols. Grassmann integration amounts to Grassmann 
differentiation. The anti-chiral covariant derivative in the chiral superspace
basis is $\bar{D}_{\dt{\a}}=-\pa/ \pa \bar{\q}^{\dt{\a}}$. The field component 
expansion of a chiral superfield $\F$ reads 
\be \label{chsf}
 \F(y,\q)= \f(y) +\sqrt{2}\q\c(y)+\q^2 M(y)~~.\ee
An anti-chiral superfield $\Bar{\F}$ in the chiral basis is given by
\be \label{antichsf}
\eqalign{
\Bar{\F}(y^{\m}-2i\q\s^{\m}\bar{\q},\bar{\q})=  ~ & ~ 
\bar{\f}(y) + \sqrt{2}\bar{\q}\bar{\c}(y) +\bar{\q}^2\bar{M}(y)  \cr
 ~ & ~ +\sqrt{2}\q\left( i\s^{\m}\pa_{\m}\bar{\c}(y)\bar{\q}^2-i\sqrt{2}\s^{\m}
\bar{\q}\pa_{\m}\bar{\f}(y)\right)+\q^2\bar{\q}^2\bo\bar{\f}(y)~,\cr}
\ee
where $\bo =\pa_{\m}\pa_{\m}$. The bars over fields serve to distinguish 
between the `left' and `right' components that are truly independent in 
Euclidean space.

The non-anticommutativity $C_{\a\b}\neq 0$ also explicitly breaks {\it half} 
of the original N=1 supersymmetry \cite{sei}. Only the chiral subalgebra 
generated by the chiral supercharges (in the chiral basis) 
$Q_{\a}=\pa/\pa\q^{\a}$ is preserved, with $\{ Q_{\a},Q_{\b} \}_*=0$, thus
defining what is now called N=1/2 supersymmetry. The use of the NAC-deformed
superspace allows one to keep N=1/2 supersymmetry manifest. The N=1/2
supersymmetry transformation laws of the chiral and anti-chiral superfield
components in eqs.~(\ref{chsf}) and (\ref{antichsf}) are as follows:
\be \label{1/2tl1}
\d\f=\sqrt{2}\ve^{\a}\c_{\a}~~,\quad \d\c_{\a}=\sqrt{2}\ve_{\a}M~,\quad
\d M =0~,\ee
and
\be \label{1/2tl2}
\d\bar{\f}=0~,\quad \d\bar{\c}^{\dt{\a}}=-i\sqrt{2}
(\tilde{\s}_{\m})^{\dt{\a}\b}\ve_{\b}\pa_{\m}\bar{\f}~,\quad
\d\bar{M}=-i\sqrt{2}\pa_{\m}\bar{\c}_{\dt{\a}}(\tilde{\s}_{\m})^{\dt{\a}\b}
\ve_{\b}~,\ee
respectively, where we have introduced the N=1/2 supersymmetry (chiral) 
parameter $\ve^{\a}$.

To the end of this section, we are going to demonstrate that, in the case of
the supersymmetric NLSM, its NAC superworldvolume gives rise to the induced
smearing or fuzzyness in the NLSM target space \cite{alv,p3}. Here we follow 
ref.~\cite{p3} where the most general four-dimensinal supersymmetric NLSM with
an arbitrary scalar potential in the NAC superspace was considered (without
any gauge fields), with the action 
\be\label{act}   S[\F,\Bar{\F}]
 = \int d^4 y \left[ \int d^2\q d^2\bar{\q}\, K(\F^i,\Bar{\F}{}^{\bar{j}})+
\int d^2\q\, W(\F^i) + \int  d^2\bar{\q}\,\Bar{W}(\Bar{\F}{}^{\bar{j}})
\right]~.\ee
This action is completely specified by the K\"ahler superpotential 
$K(\F,\Bar{\F})$, the scalar superpotential $W(\F)$, and the anti-chiral 
superpotential $\Bar{W}(\Bar{\F})$, in terms of some number $n$  of 
chiral and anti-chiral superfields, $i,\bar{j}=1,2,\ldots,n$. In Euclidean 
superspace the functions $W(\F)$ and $\Bar{W}(\Bar{\F})$ are independent upon 
each other. The NAC-deromed action is formally obtained by replacing all 
superfield profucts in  eq.~(\ref{act}) by their star products  (\ref{starp}).

The NAC-deformed extension of eq.~(\ref{act}) in four dimensions after a 
`Seiberg-Witten map' (i.e. after an explicit computation of all star products) 
was found in a closed form (i.e. in terms of finite functions) in 
refs.~\cite{p3,p1,p2}. Our four-dimensional results are in agreement with the
 results of ref.~\cite{alv} in the case of the NAC-deformed N=2 
supersymmetric two-dimensionl NLSM, after dimensional reduction to two 
dimensions.

We use the following notation valid for any function $F(\f,\bar{\f})$:
\be \label{not4}
 F_{,i_1i_2\cdots i_s \bar{p}_1\bar{p}_2\cdots \bar{p}_t}=
\fracmm{\pa^{s+t}F}{\pa\f^{i_1}\pa\f^{i_2}\cdots\f^{i_s}
\pa\bar{\f}^{\bar{p}_1}\pa\bar{\f}^{\bar{p}_2}\cdots\pa
\bar{\f}^{\bar{p}_t}}~~~~,\ee
and the Grassmann integral normalisation $ \int d^2\q\,\q^2=1$. The actual
deformation parameter, in the case of the NAC-deformed field theory 
(\ref{act}), appears to be 
\be \label{defpar}
 c=\sqrt{-\det C}~,\ee
where we have used the definition \cite{sei}
\be\label{def3}
  \det C = \fracm{1}{2}\ve_{\a\g}\ve_{\b\d}C^{\a\b}C^{\g\d}~~.\ee
As a result, unlike the case of the NAC-deformed supersymmetric gauge theories
 \cite{sei}, the NAC-deformation of the NLSM field theory (\ref{act}) appears 
to be invariant under Euclidean translations and rotations. 

A simple non-perturbative formula, describing an arbitrary NAC-deformed 
scalar superpotential $V$ depending upon a single chiral superfield $\F$, was 
found in ref.~\cite{p1},
\be \label{nacpot}
\eqalign{
 \int d^2\q\,V_*(\F)= &~ ~\fracmm{1}{2c}\left[ V(\f+cM)-V(\f-cM)\right] \cr
& -\fracmm{\c^2}{4cM}\left[ V_{,\f}(\f+cM)- V_{,\f}(\f-cM)\right]~.}\ee
The NAC-deformation in the single superfield case thus gives rise to the split
of the scalar potential, which is controlled by the auxiliary field $M$. When 
using an elementary identity
\be \label{id2}
 f(x+a)-f(x-a) =a\fracmm{\pa}{\pa x}
\int^{+1}_{-1} d\x\,f(x+\x a)~,\ee
valid for any function $f$, we can rewrite eq.~(\ref{nacpot}) to the 
equivalent form \cite{alv}
\be \label{nacpot2}
 \int d^2\q\,V_*(\F)=\fracm{1}{2}M\fracmm{\pa}{\pa\f}\int^{+1}_{-1} d\x\,
V(\f+\x cM) -\fracmm{1}{4}\c^2\fracmm{\pa^2}{\pa\f^2}\int^{+1}_{-1} d\x\,
V(\f+\x cM)~.\ee    
Similarly, in the case of several chiral superfields, one finds \cite{alv}
\be \label{nacpot3}
 \int d^2\q\,V_*(\F^I)=\fracm{1}{2}M^I\fracmm{\pa}{\pa\f^I}\Tilde{V}(\f,M)
-\fracmm{1}{4}(\c^I\c^J)\fracmm{\pa^2}{\pa\f^I\pa\f^J}\Tilde{V}(\f,M) \ee
in terms of the auxiliary pre-potential 
\be \label{nacpot4}
\Tilde{V}(\f,M)=\int^{+1}_{-1} d\x\,V(\f^I+\x cM^I)~.\ee
Hence the NAC-deformation of a generic scalar superpotential $V$ results
in its smearing or fuzziness controlled by the auxiliary fields $M^I$.

A calculation of the NAC deformed K\"ahler potential 
\be \label{nackal}
 \int d^4y\,L_{\rm kin.}\equiv\int d^4y\int d^2\q d^2\bar{\q}\, 
K(\F^i,\Bar{\F}{}^{\bar{j}})_* \ee
can be reduced to eqs.~(\ref{nacpot}) or (\ref{nacpot3}), when using a chiral
reduction in superspace, with the following result \cite{p2}:
\be \label{nackal2}
 \eqalign{
L_{\rm kin.}~ = ~& ~ \fracm{1}{2} M^iY_{,i}+
\fracm{1}{2}\pa^{\m}\bar{\f}{}^{\bar{p}}\pa_{\m}\bar{\f}{}^{\bar{q}}
Z_{,\bar{p}\bar{q}} + \fracm{1}{2}\bo\bar{\f}{}^{\bar{p}}
Z_{,\bar{p}} -\fracm{1}{4}(\c^i\c^j)Y_{,ij} \cr
 ~& ~ -\fracm{1}{2}i(\c^i\s^{\m}\bar{\c}{}^{\bar{p}})\pa_{\m}
\bar{\f}{}^{\bar{q}}Z_{,i\bar{p}\bar{q}}
-\fracm{1}{2}i(\c^i\s^{\m}\pa_{\m}\bar{\c}{}^{\bar{p}})Z_{,i\bar{p}}~,\cr}\ee
where we have introduced the (component) smeared K\"ahler pre-potential 
\be\label{smkal}
 Z(\f,\bar{\f},M)=\int^{+1}_{-1}d\x\, K^{\x} \quad {\rm with}\quad
K^{\x}\equiv K(\f^i+\x cM^i,\bar{\f}{}^{\bar{j}})~~,\ee
as well as the extra (auxiliary) pre-potential \cite{alv}
\be \label{nackal3}
 Y(\f,\bar{\f},M,\bar{M})=\bar{M}{}^{\bar{p}}Z_{,\bar{p}}
 -\fracm{1}{2}(\bar{\c}{}^{\bar{p}}\bar{\c}{}^{\bar{q}})
Z_{,\bar{p}\bar{q}} +c\int^{+1}_{-1}d\x \x\left[ 
\pa^{\m}\bar{\f}{}^{\bar{p}}\pa_{\m}\bar{\f}{}^{\bar{q}}
K^{\x}_{,\bar{p}\bar{q}} +\bo\bar{\f}{}^{\bar{p}}K^{\x}_{,\bar{p}}\right]
\ee
It is not difficult to check that eq.~(\ref{nackal2}) 
does reduce to the standard (K\"ahler) N=1 supersymmetric NLSM ({\sf cf.} 
sect.~4) in  the limit $c\to 0$. Also, in the case of a free (bilinear) 
K\"ahler potential $K=\d_{i\bar{j}}\F^i\bar{\F}^{\bar{j}}$, there is no 
deformation at all. 

The NAC-deformed scalar superpotentials $W(\F)_*$ and $\bar{W}(\bar{\F})_*$ 
imply, via eqs.~(\ref{nacpot3}) and (\ref{nacpot4}), that the following 
component terms are to be added to eq.~(\ref{nackal2}):
\be \label{nackal5}
 L_{\rm pot.}=\fracm{1}{2}M^i\Tilde{W}_{,i}-\fracm{1}{4}(\c^i\c^j)
\Tilde{W}_{,ij}+ \bar{M}^{\bar{p}}\bar{W}_{,\bar{p}}-  
\fracm{1}{2}(\bar{\c}^{\bar{p}}\bar{\c}^{\bar{q}})\bar{W}_{,\bar{p}\bar{q}}~,
\ee
where we have introduced the smeared scalar pre-potential \cite{alv}
\be \label{nacpot5}
 \Tilde{W}(\f,M)= \int^{+1}_{-1}d\x\, W(\f^i+\x cM^i)~~.\ee
The anti-chiral superpotential terms are {\it inert} under the NAC-deformation.

The $\x$-integrations in the equations above represent the smearing effects. 
However, the smearing is merely apparent in the case of a single chiral 
superfield, which gives rise to the splitting (\ref{nacpot}) only. This can 
also be directly demonstrated from eq.~(\ref{nackal2}) when using the identity
 (\ref{id2}) together with the related identity \cite{p2}
\be \label{id3}
 f(x+a) + f(x-a) =\int^{+1}_{-1}d\x\,f(x+\x a) + a\fracmm{\pa}{\pa x}
\int^{+1}_{-1}d\x\,\x f(x+\x a)~.\ee 
The single superfield case thus appears to be special, so that a sum of 
eq.~(\ref{nackal2}) and (\ref{nackal5}) can be rewritten to the bosonic 
contribution \cite{p2}
\be \label{bos}
\eqalign{
L_{\rm bos.} = & 
+\fracm{1}{2}\pa^{\m}\bar{\f}\pa_{\m}\bar{\f}\left[
K_{,\bar{\f}\bar{\f}}(\f+cM,\bar{\f})+K_{,\bar{\f}\bar{\f}}(\f-cM,\bar{\f})
\right] \cr
& + \fracm{1}{2}\bo\bar{\f}\left[
K_{,\bar{\f}}(\f+cM,\bar{\f})+K_{,\bar{\f}}(\f-cM,\bar{\f})\right] \cr
& + \fracmm{\bar{M}}{2c}\left[ K_{,\bar{\f}}(\f+cM,\bar{\f}) 
 -K_{,\bar{\f}}(\f-cM,\bar{\f}) \right] \cr
& + \fracmm{1}{2c}\left[ W(\f+cM) -W(\f-cM)\right]+\bar{M}
\fracmm{\pa\bar{W}}{\pa\bar{\f}}~~,\cr} \ee
supplemented by the following fermionic terms \cite{p2}:
\be \label{ferm}
\eqalign{
L_{\rm ferm.} = & -\fracmm{1}{4c}\bar{\c}^2\left[ 
K_{,\bar{\f}\bar{\f}}(\f+cM,\bar{\f})-K_{,\bar{\f}\bar{\f}}(\f-cM,\bar{\f})
\right] \cr
& -\fracmm{i}{2cM}(\c\s^{\m}\bar{\c})\pa_{\m}\bar{\f}\left[
K_{,\bar{\f}\bar{\f}}(\f+cM,\bar{\f})-K_{,\bar{\f}\bar{\f}}(\f-cM,\bar{\f})
\right] \cr
& -\fracmm{i}{2cM}(\c\s^{\m}\pa_{\m}\bar{\c})\left[
K_{,\bar{\f}}(\f+cM,\bar{\f})-K_{,\bar{\f}}(\f-cM,\bar{\f})
\right] \cr
& -\fracmm{\bar{M}}{4cM}\c^2 
\left[ K_{,\f\bar{\f}}(\f+cM,\bar{\f})-K_{,\f\bar{\f}}(\f-cM,\bar{\f})
\right]  \cr
& -\fracmm{1}{4M}\c^2 \pa^{\m}\bar{\f}\pa_{\m}\bar{\f} 
\left[ K_{,\f\bar{\f}\bar{\f}}(\f+cM,\bar{\f})+K_{,\f\bar{\f}\bar{\f}}
(\f-cM,\bar{\f})\right]  \cr
& +\fracmm{1}{4cM^2}\c^2 \pa^{\m}\bar{\f}\pa_{\m}\bar{\f} 
\left[ K_{,\bar{\f}\bar{\f}}(\f+cM,\bar{\f})-K_{,\bar{\f}\bar{\f}}
(\f-cM,\bar{\f})\right]  \cr
& -\fracmm{1}{4M}\c^2\bo\bar{\f}
\left[ K_{,\f\bar{\f}}(\f+cM,\bar{\f})+K_{,\f\bar{\f}}
(\f-cM,\bar{\f})\right]  \cr 
& +\fracmm{1}{4cM^2}\c^2\bo\bar{\f}
\left[ K_{,\bar{\f}}(\f+cM,\bar{\f})-K_{,\bar{\f}}
(\f-cM,\bar{\f})\right]  \cr 
& +\fracmm{1}{8cM}\c^2\bar{\c}^2
\left[ K_{,\f\bar{\f}\bar{\f}}(\f+cM,\bar{\f})-K_{,\f\bar{\f}\bar{\f}}
(\f-cM,\bar{\f})\right]  \cr\
& -\fracmm{1}{4cM}\c^2\left[ W_{,\f}(\f+cM) - W_{,\f}(\f-cM)\right]
-\fracm{1}{2}\bar{\c}^2 \bar{W}_{,\bar{\f}\bar{\f}}~.\cr}\ee

The anti-chiral auxiliary fields $\bar{M}^{\bar{p}}$ enter the action 
(\ref{nackal2}) linearly (as Lagrange multipliers), while their algebraic 
equations of motion,
\be \label{aux}
\fracm{1}{2}M^i Z_{,i\bar{p}} - \fracm{1}{4}(\c^i\c^j)Z_{,ij\bar{p}}
+\bar{W}_{,\bar{p}}=0~, \ee
are the non-linear set of equations on the auxiliary fields 
$M^i=M^i(\f,\bar{\f})$.~\footnote{Equation (\ref{aux}) is not a linear system 
because the function $Z$ is $M$-dependent.} As a result, 
the bosonic scalar potential in components is given by \cite{p2} 
\be \label{aux2}
V_{\rm scalar}(\f,\bar{\f}) = 
\left. \fracm{1}{2}M^i\Tilde{W}_{,i}\right|_{M=M(\f,\bar{\f})}~~.\ee
	
Some comments are in order.

The NAC-deformation just described is only possible in Euclidean superspace 
where the chiral and anti-chiral spinors are truly independent.

The NAC-deformed NLSM is completely specified by a K\"ahler function
$K(\F,\bar{\F})$, a chiral function $W(\F)$, an anti-chiral function 
$\bar{W}(\bar{\F})$ and a constant deformation parameter $c$. As a matter of
fact, we didn't really use the constancy of $c$, so our results are valid 
even for any coordinate-dependent NAC deformation with $c(y)$. 

Solving for the auxiliary fields in eq.~(\ref{nackal2}) represents not only a 
technical but also a conceptual problem because of the smearing effects 
described by the $\x$-integrations. To bring the kinetic terms in 
eqs.~(\ref{nackal2}) or (\ref{bos}) to the standard NLSM form (i.e. without 
the second order derivatives), one  has to integrate by parts that 
leads to the appearance of the derivatives of the auxiliary fields. 
This implies that one has to solve eq.~(\ref{aux}) {\it before} integration by
 parts.  Let $M^i=M^i(\f,\bar{\f})$ be a solution to eq.~(\ref{aux}), and 
let's ignore fermions for simplicity 
$(\c^i_{\a}=\bar{\c}_{\dt{\a}}^{\bar{p}}=0)$. Substituting the auxiliary field
solution back to the Lagrangian (\ref{nackal2}) and integrating by parts yield
\be \label{nackal6}
\eqalign{
L_{\rm kin.} (\f,\bar{\f})= & 
-\fracm{1}{2}(\pa_{\m}\bar{\f}^{\bar{p}}\pa_{\m}\f^q)\int^{+1}_{-1}d\x\left[
K^{\x}_{,\bar{p}q}+2c\x M^i_{,q}K^{\x}_{,\bar{p}i}+c\x M^iK^{\x}_{,\bar{p}iq}
\right. \cr
 &\left. +c^2\x^2M^iK^{\x}_{,\bar{p}ij}M^j_{,q}\right] \cr
& -\fracm{1}{2}(\pa_{\m}\bar{\f}^{\bar{p}}\pa_{\m}\bar{\f}^{\bar{q}})
\int^{+1}_{-1}d\x\left[ 2c\x K^{\x}_{,\bar{p}i}M^i_{,\bar{q}}+c^2\x^2M^i
K^{\x}_{,\bar{p}ij}M^j_{,\bar{q}}\right]~~.\cr}\ee
It is now apparent that the NAC-deformation does not preserve the original 
K\"ahler geometry of eq.~(\ref{act}), though the absence of $(\pa_{\m}\f)^2$ 
terms and the particular structure of various contributions to 
eq.~(\ref{nackal6}) are quite remarkable. The action (\ref{nackal6}) takes the
 form of a generic NLSM, being merely dependent upon mixed derivatives of the 
K\"ahler function, so that the original K\"ahler gauge invariance of 
eq.~(\ref{act}), 
\be\label{kgau}
 K(\f, \bar{\f})\to K(\f, \bar{\f})+f(\f) + \bar{f}(\bar{\f})~,\ee
with arbitrary gauge functions $f(\f)$ and $\bar{f}(\bar{\f})$ is preserved.
See ref.~\cite{p1} for more discussion about elimination of the auxiliary
fields.

As a result, the NAC deformation of the NLSM (\ref{act}) amounts to the
non-K\"ahlerian and non-Hermitian deformation of the original K\"ahlerian and 
Hermitian structures, which is controlled by the auxiliary field solution to 
eq.~(\ref{aux}). In the case of a single chiral superfield, the deformed NLSM 
metric can be read off from the following kinetic terms \cite{p1}:
\be\label{kin1}\eqalign{
L_{\rm kin.} (\f,\bar{\f})= &  
-\fracm{1}{2}(\pa_{\m}\bar{\f}\pa_{\m}\f)\fracmm{\pa}{\pa\f}
\fracmm{\pa}{\pa\bar{\f}} 
\left[ K(\f+cM(\f,\bar{\f}),\bar{\f}) +K(\f-cM(\f,\bar{\f}),\bar{\f})\right]\cr
& +\fracm{1}{2}(\pa_{\m}\bar{\f}\pa_{\m}\f)\fracmm{\pa}{\pa\f}
\left[ cK_{,\f}(\f+cM,\bar{\f}) -cK_{,\f}(\f-cM,\bar{\f})
\right]\fracmm{\pa M(\f,\bar{\f})}{\pa\bar{\f}} \cr
 & -\fracm{1}{2}(\pa_{\m}\bar{\f}\pa_{\m}\bar{\f}) \left[ 
cK_{,\f\bar{\f}}(\f+cM(\f,\bar{\f}),\bar{\f})-cK_{,\f\bar{\f}}(\f-
cM(\f,\bar{\f}),\bar{\f})\right] \cr
& \times \fracmm{\pa M(\f,\bar{\f})}{\pa\bar{\f}}~~.\cr}\ee
In the case of several superfields, the deformed NLSM can be read 
off from eq.~(\ref{nackal6}), when assuming all the $\x$-integrations to be 
performed with the auxiliary fields considered as spectators.

We thus find a new (NAC) mechanism of deformation of complex geometry in the 
supersymmetric NLSM target space, by using a non-vanishing anti-holomorphic 
scalar potential $\Bar{W}(\Bar{\F})$, because elimination of the auxiliary 
fields $M$ and $\bar{M}$ via their algebraic equations of motion in the NAC
deformed NLSM results in the deformed {\it bosonic} K\"ahler potential 
depending upon $C$ and $\Bar{W}'(\Bar{\F})$. This feature is specific to the 
NAC deformation, because a scalar potential does not affect a K\"ahler 
potential in the usual (undeformed) NLSM. 

\section{Example: NAC-deformed $CP(1)$ model}

Let's consider the simplest non-trivial example provided by a
four-dimensional supersymmetric $CP(1)$ NLSM with the (undeformed) K\"ahler, 
Hermitian and symmetric target space characterized by the K\"ahler potential
\be\label{cp1}
K(\f,\bar{\f})=\a\, \ln (1 +\k^{-2}\f\bar{\f})~~,\ee
where two dimensional constants, $\a$ and $\k$, have been introduced, and 
with an arbitrary anti-holomorphic scalar superpotential $\Bar{W}(\Bar{\F})$. 
An explicit solution to the auxiliary field equation (\ref{aux}) in this case 
reads \cite{p1}
\be \label{ms}
M =\fracmm{\a -\sqrt{\a^2+\left(2c\bar{\f}(1+\k^{-2}\f\bar{\f})
\bar{W}_{,\bar{\f}}\right)^2}}{2c^2\k^{-2}\bar{\f}^2\bar{W}_{,\bar{\f}}}~~,
\ee
where we have used the notation $\bar{W}_{,\bar{\f}}=\pa\bar{W}/\pa\bar{\f}$. 
A straightforward calculation yields the following deformed NLSM kinetic 
terms \cite{p1}:
\be \label{cp1k}
L_{\rm kin.}= -g\low{\f\f}\pa_{\m}\f\pa_{\m}\f
  -2g\low{\f\bar{\f}}\pa_{\m}\f \pa_{\m}\bar{\f}
- g\low{\bar{\f}\bar{\f}}\pa_{\m}\bar{\f}\pa_{\m}\bar{\f}~~,\ee
where
\be \label{cp1me}
\eqalign{
g\low{\f\bar{\f}} =~& 
\fracmm{-\a\k^{-2}c^2\bar{\f}^2(\bar{W}_{,\bar{\f}})^2}{
\left(-\a+\sqrt{\a^2+\left(2c\bar{\f}(1+\k^{-2}\f\bar{\f})
\bar{W}_{,\bar{\f}}\right)^2}\right)
\sqrt{\a^2+\left(2c\bar{\f}(1+\k^{-2}\f\bar{\f})\bar{W}_{,\bar{\f}}\right)^2}},
\cr
g\low{\f\f} = & ~0~,\cr
g\low{\bar{\f}\bar{\f}} =~ & 
\fracmm{-2\a^{-1}c^2(1+\k^{-2}\f\bar{\f})\bar{W}_{,\bar{\f}}}{\left(
\a-\sqrt{\a^2+\left(2c\bar{\f}(1+\k^{-2}\f\bar{\f})
\bar{W}_{,\bar{\f}}\right)^2}\right)
\sqrt{\a^2+\left(2c\bar{\f}(1+\k^{-2}\f\bar{\f})
\bar{W}_{,\bar{\f}}\right)^2}} ~\times \cr
& \times \left[ 4c^2\bar{\f}^2(\bar{W}_{,\bar{\f}})^3(1+\k^{-2}\f\bar{\f}) 
\right.  \cr
& \left. +~ \a\left( \a-\sqrt{\a^2+\left(2c\bar{\f}(1+\k^{-2}\f\bar{\f})
\bar{W}_{,\bar{\f}}\right)^2}\right) (2\bar{W}_{,\bar{\f}}+\bar{\f}
\bar{W}_{,\bar{\f}\bar{\f}})\right].\cr}\ee

It is worth noticing that $\det g =-(g\low{\f\bar{\f}})^2$. The most apparent
feature $g\low{\f\f}=0$ is also valid in the case of a generic NAC-deformed 
NLSM (in the given parametrization). 

\section{Conclusion}

Our approach to the NAC-deformed NLSM is very general. The NAC deformation 
(i.e. smearing or fuzziness) of the NLSM K\"ahler potential is controlled by 
the auxiliary fields $M^i$ entering the deformed K\"ahler potential in the 
highly non-linear way. Both locality and Euclidean invariance are preserved, 
while no higher derivatives appear in the deformed NLSM action.

One should distinguish between the NAC-deformation and N=1/2 supersymmetry.
Though the NAC-deformation we considered is N=1/2 supersymmetric, the former
is {\it stronger} than the latter. When requiring merely N=1/2 supersymmetry 
of a four-dimensional NLSM, it would give rise to much weaker restrictions on 
the NLSM target space. 

It is still the open question how to describe the NAC deformation of the NLSM 
metric in purely geometrical terms.

This investigation was supported in part by the Japanese Society for Promotion 
of Science (JSPS). I am grateful to K. Ito, O. Lechtenfeld,  P. Bowknegt and
 S. Watamura for useful discussions during the workshop.
\vglue.2in


\begin{thebibliography}{99}

\bibitem{nlsm} Ketov S.V.: Quantum Non-Linear Sigma Models. Springer-Verlag,
Berlin Heidelberg New-York (2000)
\bibitem{sev4} Spindel, P., Sevrin, A., Troost, W., van Proeyen, A.:
Extended Supersymmetric Sigma Models on Group Manifolds. Nucl. Phys.
{\bf B308}, 662--698 (1988), and {\bf B311}, 465--492 (1988)
\bibitem{sevrin} Sevrin, A., Troost J.: Off-shell Formulation of N=2
Non-linear Sigma-models. Nucl. Phys. {\bf B492}, 623--646 (1997)
\bibitem{cdg} Huybrechts D.: Complex Geometry. Springer-Verlag, 
Berlin Heidelberg New-York (2004)
\bibitem{ghr} Gates, S.J. Jr., Hull, C., Ro\v{c}ek, M.:  
Twisted Multiplets and New Supersymmetric Non-linear Sigma Models.
 Nucl. Phys. {\bf B248}, 157--186 (1984) 
\bibitem{bus} Buscher, T., Lindstr\"om, U.,  Ro\v{c}ek, M.:
New Supersymmetric Sigma-models with Wess-Zumino Terms. Phys. Lett. {\bf B202},
 94--98 (1988)
\bibitem{gk} Gates, S.J. Jr., Ketov, S.V.: No N=4 Strings on Wolf Spaces.
Phys. Rev. {\bf D52}, 2278--2293 (1995)
\bibitem{hss} Galperin, A., Ivanov, E., Ogievetsky, V., Sokatchev, E.: 
Harmonic Superspace. Cambridge University Press (2001) 
\bibitem{bs} Brink, L., Schwarz J.: Quantum Superspace. Phys.Lett. {\bf B100},
310--312 (1981)
\bibitem{kpt} Klemm, D., Penati, S., Tamassia, L.: Non(anti)commutative
Superspace. Classical and Quantum Grav. {\bf 20}, 2905--2916 (2003)   
\bibitem{ov} Ooguri, H., Vafa C.: Gravity induced C Deformation. 
Adv. Theor. Math. Phys. {\bf 7}, 405--417 (2004)
\bibitem{kgn} Gates, S.J. Jr., Ketov, S.V., Nishino, H.: Self-dual 
Supersymmetry and Supergravity in Atiyah-Ward Space-time. Nucl. Phys. 
{\bf B716}, 149--210 (1993)
\bibitem{sei} Seiberg, N.: Noncommutative Superspace, N=1/2 Supersymmetry and
String Theory. JHEP {\bf 0306}, 010 (2003)
\bibitem{alv} Alvarez-Gaume, L., Vazquez-Mozo, M.: On Non-anti-commutative N=2
Sigma-models in Two Dimensions. JHEP {\bf 0504}, 007 (2005)
\bibitem{p3} Hatanaka, T., Ketov, S.V., Kobayashi, Y., Sasaki, S.: 
N=1/2 Supersymmetric Four-dimensional Non-linear $\s$-models from
Non-anti-commutative Superspace. Nucl. Phys. {\bf B726}, 481--493 (2005)
\bibitem{p1} Hatanaka, T., Ketov, S.V., Kobayashi, Y., Sasaki, S.: 
Non-anti-commutative Deformation of Effective Potentials in Supersymmetric
Gauge Theories. Nucl. Phys. {\bf B716}, 88--104 (2005)
\bibitem{p2} Hatanaka, T., Ketov, S.V., Sasaki, S.: Summing up 
Non-anti-commutative K\"ahler Potential. Phys. Lett. {\bf B619}, 352--358 
(2005).
\end{thebibliography}
\end{document}